\documentclass[structabstract]{aa}

\usepackage{lscape}
\usepackage{txfonts}
\usepackage{graphicx}
\usepackage{rotating}
\usepackage{natbib}
\usepackage{aalongtable}
\usepackage{gensymb}
\usepackage{url}
\bibpunct[; ]{(}{)}{,}{a}{}{;} 

\begin{document}

\title{VLT/MUSE discovers a jet from the evolved B[e] star MWC~137\thanks{Based on observations collected at the European Southern Observatory, Chile: Prog.ID 094.D-0215(A).}\\ 
} 

\titlerunning{A jet emerging from the evolved B[e] star MWC~137}


\author{A. Mehner\inst{1}  
\and W.J. de Wit\inst{1}
\and J.H. Groh\inst{2}
\and R.D. Oudmaijer\inst{3}
\and D. Baade\inst{4}
\and T. Rivinius\inst{1}
\and F. Selman\inst{1}
\and H.M.J. Boffin\inst{1}
\and C. Martayan\inst{1}
}

 \institute{ESO -- European Organisation for Astronomical Research in the Southern Hemisphere, Alonso de Cordova 3107, Vitacura, Santiago de Chile, Chile
 \and Geneva Observatory, Geneva University, Chemin des Maillettes 51, 1290 Versoix, Switzerland
 \and School of Physics and Astronomy, The University of Leeds, Leeds, LS2 9JT, UK
  \and ESO -- European Organisation for Astronomical Research in the Southern Hemisphere, Karl-Schwarzschild-Stra{\ss}e 2,  85748 Garching, Germany
} 

\abstract {}{Not all stars exhibiting the optical spectral characteristics of B[e] stars share the same evolutionary stage. The Galactic B[e] star MWC~137 is a prime example of an object with uncertain classification, with previous work suggesting pre- and post-main sequence classification. Our goal is to settle this debate and provide reliable evolutionary classification.}{Integral field spectrograph observations with the Very Large Telescope Multi Unit Spectroscopic Explorer (VLT MUSE) of the cluster SH~2-266 are used to analyze the nature of MWC~137.}{A collimated outflow is discovered that is geometrically centered on MWC~137. The central position of MWC~137 in the cluster SH~2-266 within the larger nebula suggests strongly that it is a member of this cluster and that it is both at the origin of the nebula and the newly discovered jet. 
Comparison of the color-magnitude diagram of the brightest cluster stars with stellar evolutionary models results in a distance of about $5.2\pm1.4$~kpc. We estimate that the cluster is at least 3~Myr old. 
The jet originates from MWC~137 at a position angle of 18--20\degree. The jet extends over 66\arcsec\  (1.7~pc) projected on the plane of the sky, shows several knots, has electron densities of about $10^3$~cm$^{-1}$, and projected velocities of up to $\pm450$~km~s$^{-1}$.
From the Balmer emission line decrement of the diffuse intracluster nebulosity we determine  $E(B$--$V)=1.4$~mag for the inner 1\arcmin\ cluster region. The spectral energy distribution of the brightest cluster stars yield a slightly lower extinction of $E(B$--$V) \sim 1.2$~mag for the inner region and $E(B$--$V) \sim 0.4$--$0.8$~mag for the outer region. The extinction towards MWC~137 is estimated to be $E(B$--$V) \sim 1.8$~mag ($A_V \sim 5.6$~mag).}{Our analysis of the optical and near-infrared color-magnitude and color-color diagrams suggests a post-main sequence stage of MWC~137. The existence of a jet in this object implies the presence of an accretion disk. 
Several possibilities for MWC~137's nature and the origin of its jet are discussed, e.g., the presence of a companion and a merger event.}

\keywords{Stars: individual: MWC~137 -- circumstellar matter -- Stars: distances -- Stars: emission-line, Be -- Stars: evolution -- Stars: jets -- Stars: massive -- Stars: mass-loss -- Stars: winds, outflows}

\maketitle

\section{Introduction}  
\label{intro}

\begin{figure*}[!]
\centering
\begin{minipage}{.5\textwidth}
  \centering
  \includegraphics[width=.81\linewidth]{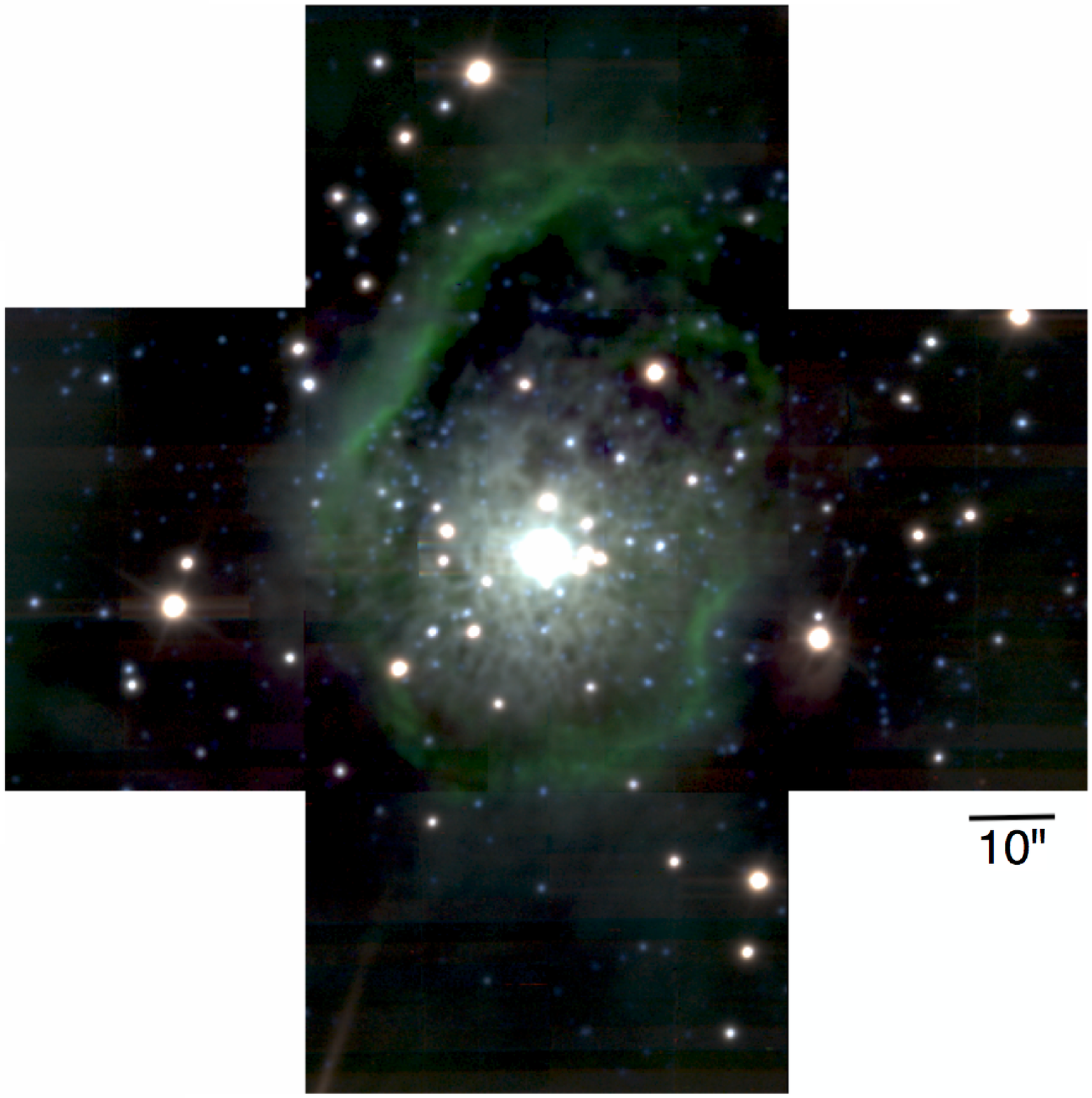}
\end{minipage}%
\begin{minipage}{.5\textwidth}
  \centering
  \includegraphics[width=1\linewidth]{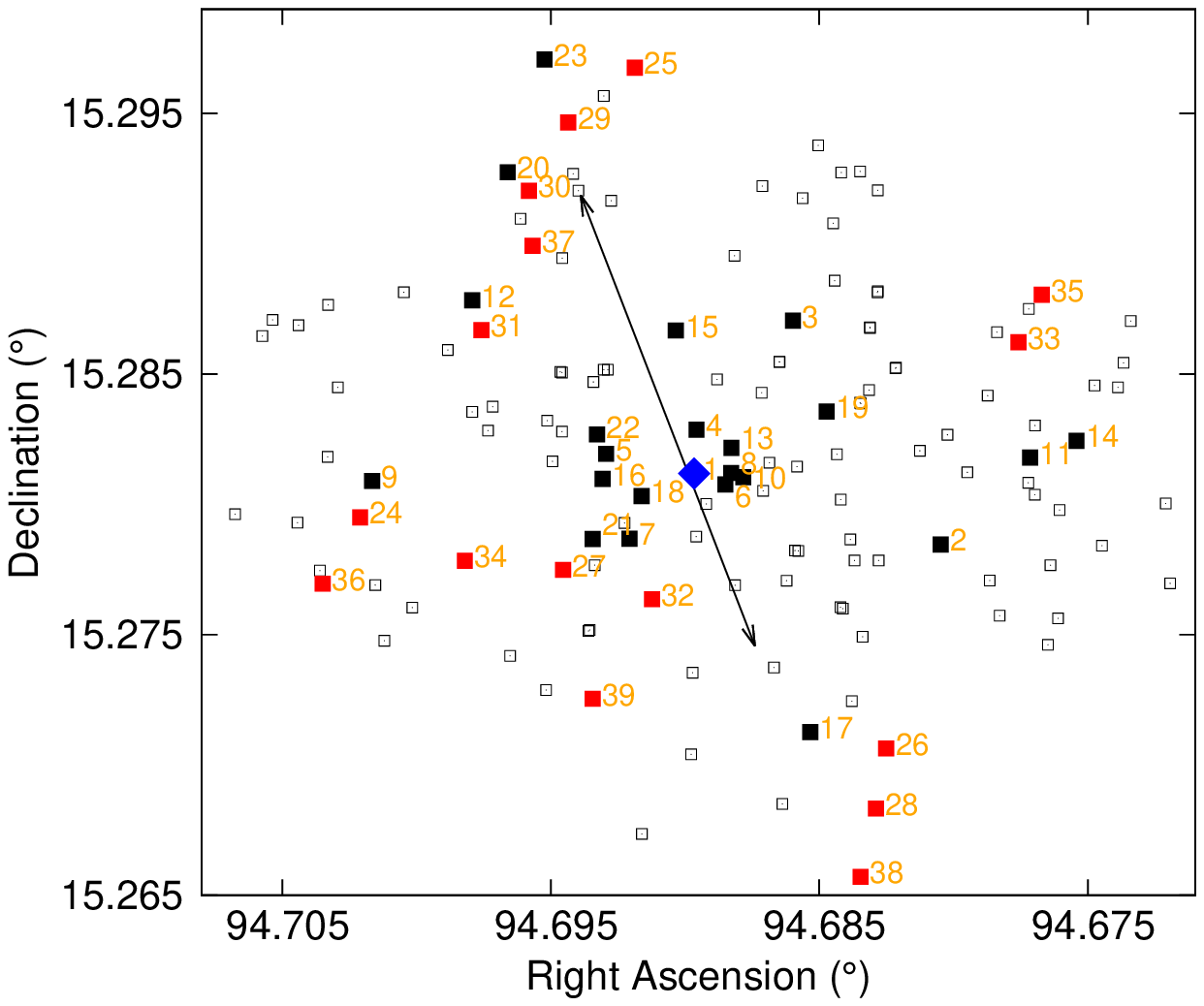}
\end{minipage}
  \caption{Left: Composite $VRI$ color image ($1.7\arcmin \times 1.7\arcmin$) of the five MUSE fields covering the cluster SH~2-266, the central star MWC~137, and its circumstellar nebula. Right: Spatial distribution of the brightest cluster stars (blue diamond in the center: MWC~137; black squares: A- and B-type stars; red squares: F- and G-type stars; black open squares: unclassified stars; arrow: jet extension, see Section \ref{jet}). The stars for which a spectral type was obtained are indicated with a number, consistent with Table \ref{table:spectraltype}.}
\label{fig:fov}
\end{figure*}

The evolutionary phase of B[e] stars is difficult to establish, because the classification criteria are met by stars of different masses and evolutionary stages \citep{1998A&A...340..117L,1998ASSL..233....1Z,2006ASPC..355.....K}. 
The B[e] designation refers to physical and spectral diagnostics, i.e., the presence of forbidden emission lines and dust emission. These characteristics are observed in stars of different evolutionary phases: supergiant B[e] stars (sgB[e]), pre-main sequence Herbig  AeB[e] stars, compact planetary nebulae around B stars, symbiotic B[e] stars, and a large group of unclassified B[e] stars, whose evolutionary phases are unknown \citep{2006ASPC..355..371L}. The unclassified B[e] stars have been proposed to form the FS~CMa class, whose properties have been suggested to be due to mass exchange in binary systems \citep{2007ApJ...667..497M}.
One of the main problems in the evolutionary classification of these objects is their unknown distance and thus their unknown luminosity.

For instance, the classification of the Galactic B[e] star MWC~137\footnote{Another common identifier is V$^*$~V1308~Ori.} includes pre-main sequence phases based on its emission-line spectrum \citep{1992ApJ...397..613H,1992ApJ...398..254B,1994A&AS..104..315T}, and post-main sequence phases based on its high luminosity, analysis of its stellar and nebular spectra, and its radio emission \citep{1972ApJ...174..401H,1984A&AS...55..109F,1998MNRAS.298..185E}. The range of distances to MWC~137 discussed in the literature is large ($d= 1-13$~kpc; e.g., \citealt{1992ApJ...397..613H,1984Ap&SS.107...19A,1984ApJ...279..125F,1988A&A...191..323W}). Its luminosity and mass are therefore extremely uncertain.

Concerning its evolutionary stage, MWC~137's large circumstellar nebula shows no enhanced nitrogen and helium abundance. \citet{1998MNRAS.298..185E} argued that the morphology and other properties of the nebula suggest that it was produced by the interaction of stellar winds with the ambient interstellar medium or unprocessed ejected matter. \citet{2008A&A...477..193M} also suggested that the nebula is unevolved, swept-up material.

Recent studies found evidence for an evolved, post-main sequence, nature of MWC~137 based on the $^{12}$C/$^{13}$C ratio in CO emitting circumstellar material close to the star \citep{2013A&A...558A..17O,2015AJ....149...13M}. They argued that the enrichment of $^{13}$C in a circumstellar ring of CO gas excludes a classification as pre-main sequence Herbig AeBe (HAeBe) star and that MWC~137 is most likely in an extremely short-lived phase evolving from a B[e] supergiant to a blue supergiant with a bipolar ring nebula. However, the $^{12}$C/$^{13}$C ratio is still compatible with that of a protoplanetary nebula and a main sequence star with a mass $M> 9~M_{\odot}$, because the $^{12}$C/$^{13}$C ratio drops for such stars already on the main sequence \citep{2009A&A...494..253K}. 
The observed value of the $^{12}$C/$^{13}$C ratio is indicative, but by itself it does not provide strong evidence for the object's evolutionary stage.

We investigate the cluster SH~2-266 to establish MWC~137's membership and constrain the star's evolutionary stage using data obtained with the new Very Large Telescope Multi Unit Spectroscopic Explorer (VLT MUSE; \citealt{2010SPIE.7735E..08B}). While MWC~137 is generally associated with this cluster \citep{1999A&A...342..515T}, no detailed study of the cluster properties exist.
In Section \ref{data} we describe the observations. In Section \ref{results} we present our results regarding the cluster and MWC~137. We analyze a collimated outflow, which we discovered in the MUSE data and which probably originates from MWC~137. In Section \ref{discussion} we discuss the potential origin of this jet with respect to the evolutionary phase of MWC~137. Section \ref{conclusion} gives a short conclusion.

\section{Data and analysis} 
\label{data}

The cluster SH~2-266 with the B[e] star MWC~137 in the center was observed with VLT MUSE on 2014-11-07 and 2014-12-01. 
MUSE is a spectrograph composed of 24 integral field units (IFUs) that sample a $1\arcmin \times 1\arcmin$ field of view  \citep{2010SPIE.7735E..08B}. The instrument covers almost the full optical domain (4800--9300~\AA) with a spectral resolving power of $R\sim3\,000$. The spatial resolution for MUSE's currently offered wide field mode is 0\farcs3--0\farcs4 (the pixel scale is 0\farcs2 per pixel). We obtained seeing-limited spatial resolutions between 0\farcs6 and 0\farcs7.

Three exposures of 2~s, 20~s, and 300~s were obtained with MWC~137 in the center of the field of view of MUSE. Additional 600~s exposures were acquired with offsets of 35\arcsec\ in each of the cardinal directions (see Figure \ref{fig:fov}). The limiting point source magnitude in our data is $V \approx 24$~mag.
MUSE provides the first complete and accurate set of optical line ratios and line centroids at each position of the cluster SH~2-266 and the large-scale nebula around MWC~137.

The data were reduced using version 1.1.0 of the MUSE standard pipeline \citep{2012SPIE.8451E..0BW}.
Bias, arc, and flat-field master calibrations were created using a set of standard calibration exposures. Bias images were subtracted from each science frame. The science frames were flat-fielded
using the master flat-field and renormalized using a flat-field taken right after the science to account for the temperature variations in the illumination pattern of the slices. An additional flat-field correction was performed using the twilight sky exposures to correct for the difference between sky and calibration unit illumination. A geometrical calibration and the wavelength calibration solution were used to transform the detector coordinate positions to wavelengths and spatial coordinates. The astrometric solution, flux calibration, and telluric correction were then applied. Sky subtraction was performed only in the post-reduction analysis of the individual stellar spectra. A data cube was produced from each pixel-table using a three-dimensional drizzle interpolation process, which also rejects cosmic rays. All data cubes were sampled to $0\farcs2\times0\farcs2\times1.25~\AA$. 

The MUSE standard pipeline also delivers field-of-view images in different filters when the filter transmission curves (in our case Johnson V, Cousins R, and Cousins I) are provided, see Figure \ref{fig:fov} for a composite $VRI$ image. We used the Aperture Photometry Tool v.2.4.7\footnote{\url{http://www.aperturephotometry.org/aptool/}} to determine the $VRI$ magnitudes of the brightest cluster members. 
Our MUSE data do not provide absolute magnitudes and we thus used $V=11.95$~mag for MWC~137 as a reference \citep{2003AJ....125.2531R}. We found no $I$-band magnitude reported in the literature. We therefore compare the $V$--$I$ colors of the stars for which we determined the spectral type and extinction  (Section \ref{spt}) to the $V$--$I$ colors reported in \citet{2013ApJS..208....9P} to ensure our values are consistent with literature values.
QFitsView\footnote{\url{http://www.mpe.mpg.de/~ott/dpuser/qfitsview.html}} was used to extract the stellar spectra using circular apertures from which an annulus mean sky/nebular spectrum was subtracted. For the analysis of the strengths of the diffuse interstellar bands (DIBs), discussed in Section \ref{spt}, spectra were not corrected for the sky/nebula component.

\section{Results}
\label{results}

\subsection{The extinction to SH~2-266 from the Balmer decrement of the diffuse intracluster gas}
\label{balmer}

\begin{figure}
\centering 
\includegraphics[width=0.385\textwidth]{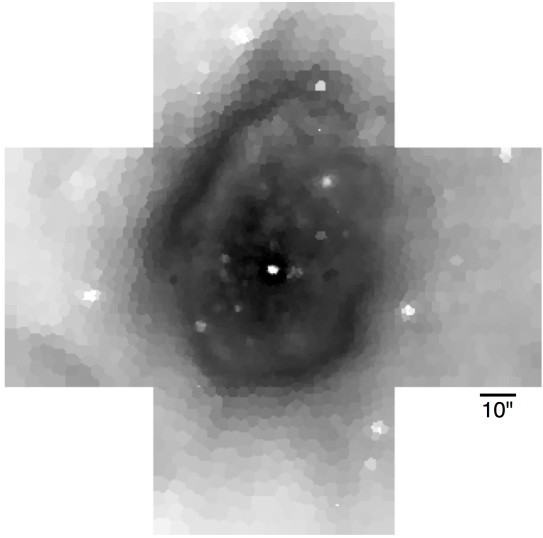}
    \caption{Voronoi diagram of $E(B$--$V)$ based on the extended nebular H$\alpha$/H$\beta$ ratio, binned to a signal to noise ratio of 200 per polygon. The color coding is linear between $E(B$--$V) = 0$~mag (white) and  $E(B$--$V) = 1.8$~mag (black). The extinction is highest close to MWC~137 and along the rim of the nebula, tracing the dust. It decreases towards the outskirts of the cluster. The white spots are stars and MWC~137 is saturated. This diagram serves for illustration purposes. For absolute extinction values, individual spectra were examined, see Section \ref{spt}. \label{fig:halpha_Hbeta}}
\end{figure}

\begin{figure}
\centering 
\resizebox{\hsize}{!}{\includegraphics{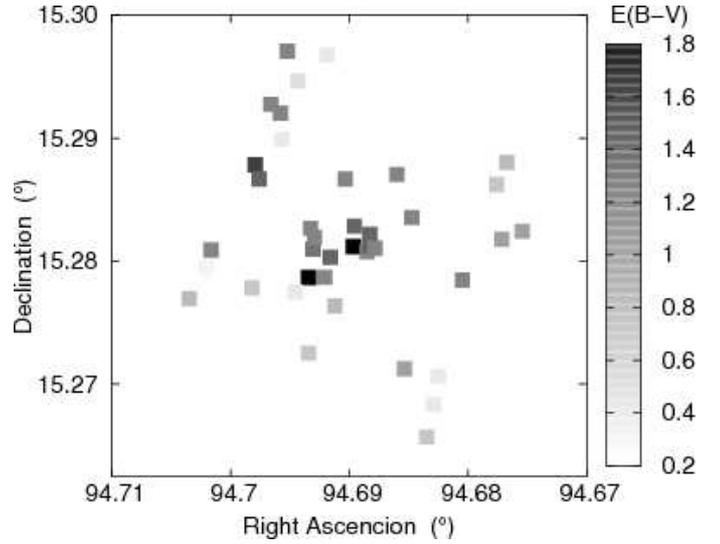}}
    \caption{Extinction map towards the stars classified in Section \ref{spt}. The extinction was determined based on the comparison of their spectral energy distributions to UVES POP and Indo-US template spectra. \label{fig:extinction_SpT}}
\end{figure}

\begin{figure*}
\centering
\begin{minipage}{.5\textwidth}
  \centering
\includegraphics[width=1\linewidth]{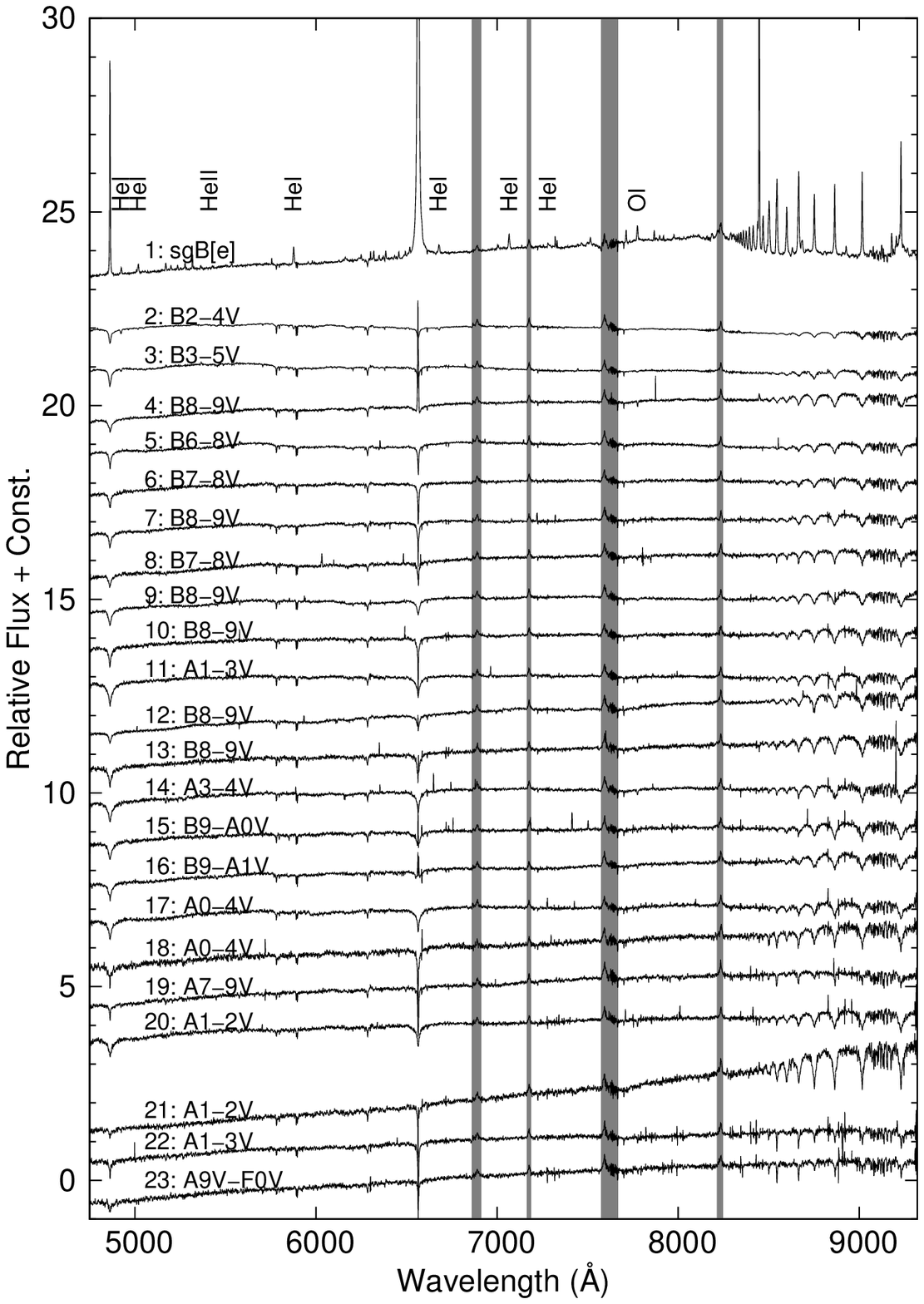}
\end{minipage}%
\begin{minipage}{.5\textwidth}
\centering 
\includegraphics[width=1\linewidth]{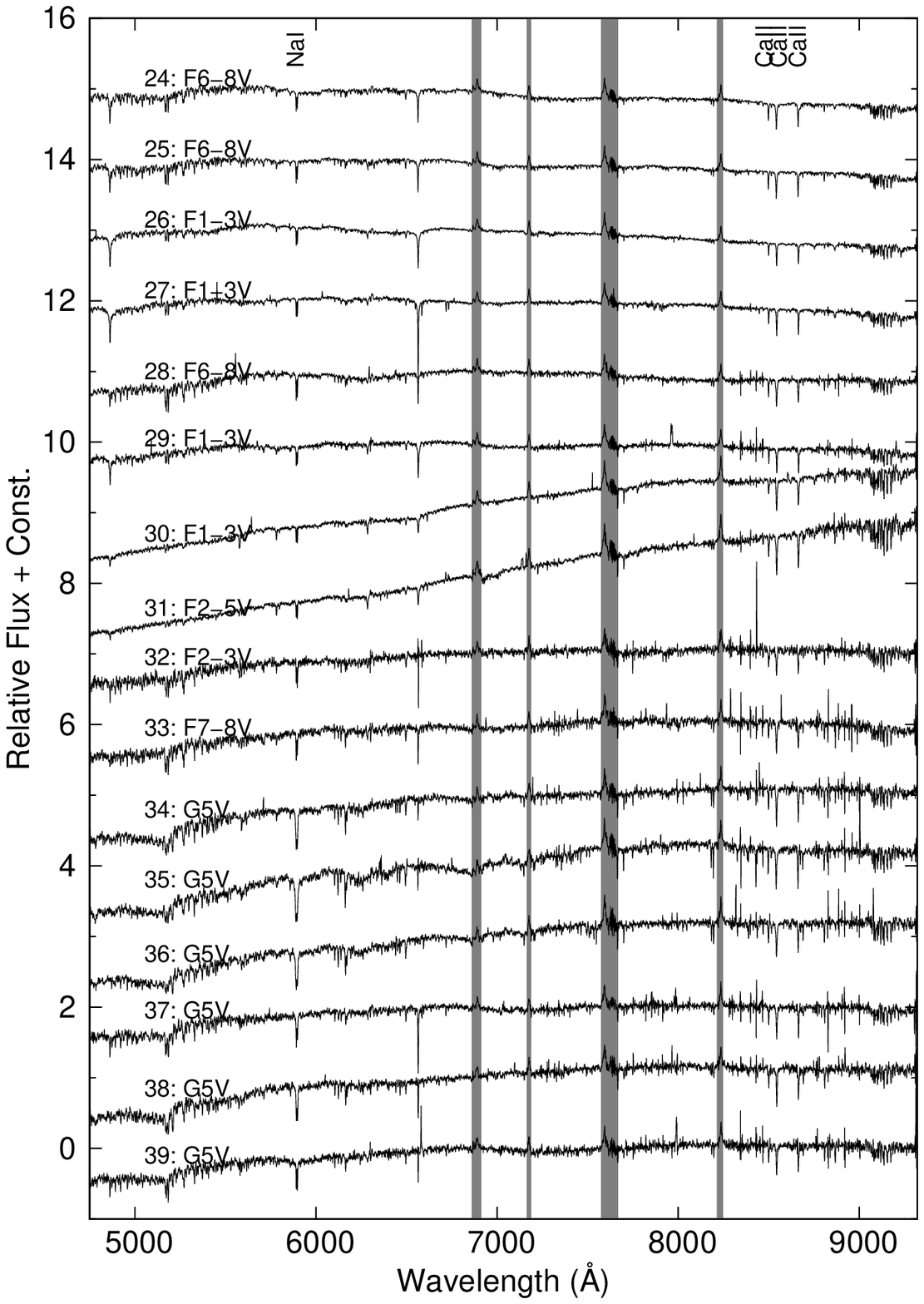}
     \end{minipage}
    \caption{Spectra of the brightest cluster stars extracted from the MUSE data cube. The spectrum \#1 is of MWC~137. Spectra \#2 to \#23 are B- and A-type stars in the cluster. Their spectral classification, indicated next to each spectrum, was based primarily on the H$\alpha$, H$\beta$, and \ion{O}{I} 7774 equivalent widths. Spectra \#24 to \#39 are F- and G-type stars. Their classification is based primarily on the \ion{Ca}{II} 8542 and \ion{Na}{I} 5889 equivalent widths. The numbering is consistent with Table \ref{table:spectraltype} and Figure \ref{fig:fov}. Grey areas block regions of the spectra with strong residuals from the telluric bands correction.}
         \label{fig:spectra}
    \end{figure*}

We used the Balmer emission line decrement across the inner 1\arcmin$\times$1\arcmin\ field of SH~2-266 to determine the extinction towards the cluster and thus a lower limit of the extinction towards MWC~137 (assuming the presence of additional circumstellar extinction). 
H$\alpha$ and H$\beta$ depend only weakly on the electron density and temperature. The comparison of the observed line intensity ratio (H$\alpha$/H$\beta)_{obs}$ to the theoretical recombination ratio (H$\alpha$/H$\beta)_{int}$ yields thus a measure of extinction.  

Case B  hydrogen recombination is typically assumed for determining the intrinsic line ratio of \ion{H}{II} regions (the nebula is optically thick to photons of the Lyman series but optically thin to Ly$\alpha$ and photons of the higher hydrogen series). We used the [\ion{S}{II}] 6716/6731 ratio and the [\ion{N}{II}] (6548+6583)/5755 ratio \citep{2006agna.book.....O} to estimate an average electron density of $n_{e}\sim3 \times 10^2$~cm$^{-3}$ and a temperature of $T\sim10\,000$~K across the nebula. The expected intrinsic (H$\alpha$/H$\beta)_{int}$ line ratio for an electron density of $n_{e}=10^2$~cm$^{-3}$ and temperature of $T=10\,000$~K is 2.863 \citep{2006agna.book.....O}.
This ratio varies by only 0.057 for electron densities $n_{e}=10^{2}$~cm$^{-3}$ to $10^{6}$~cm$^{-3}$ and by 0.294 for temperatures between 5\,000~K and 20\,000~K.


The color excess $E(B$--$V)$ due to dust attenuation can be expressed as
\begin{eqnarray}
E(B{-}V) = \frac{2.5}{\kappa(H\beta){-}\kappa(H\alpha)} \log_{10} \left[ \frac{(H\alpha/H\beta)_{obs}}{(H\alpha/H\beta)_{int}} \right]
\end{eqnarray}
(see, e.g., \citealt{2013AJ....145...47M} for a derivation of this equation).
Given that in reality the reddening law is particular for any direction in the sky, the functional form of the attenuation curve is a matter of choice. We used the Galactic reddening law of \citet{1999PASP..111...63F} to derive $\kappa(\lambda) = 21\,896/\lambda$--$0.954$ (where $\lambda$ is in {\AA}ngstr\"{o}m units) for the optical wavelength region covered by MUSE. This leads to $\kappa(H\alpha)=2.38$ and $\kappa(H\beta)=3.55$ and thus to
\begin{eqnarray}
E(B{-}V) = 2.137 \log_{10} \left[ \frac{(H\alpha/H\beta)_{obs}}{2.863} \right].
\end{eqnarray}
The difference in the intensity of the lines as a result of the use of different extinction laws is not significant compared to the uncertainties associated with our measurements.  

We determined the Balmer decrement (H$\alpha$/H$\beta)_{obs}$ across the inner 1\arcmin$\times$1\arcmin\ region of the cluster SH~2-266 with a grid size of 10\arcsec. Line intensities were measured by integrating all of the flux in a line between two suitably adopted wavelength ranges and above a local continuum. We found (H$\alpha$/H$\beta)_{obs} = 12.98 \pm 2.27$ (the error is the standard deviation of all measured values), which results in $E(B$--$V) = 1.40\pm0.47$~mag. 
We used the reddening relations from \citet{1996A&A...314..108M} to determine $E(V$--$I)=0.8 \times E(B$--$V)=1.12\pm0.38$~mag.
Adopting a ratio between total and selective extinction of $R_{V}=A_{V}/E(B$--$V)=3.1$ \citep{1989ApJ...345..245C}, we found $A_{V}=4.35\pm1.44$~mag towards the inner 1\arcmin$\times$1\arcmin\ cluster region. 

Figure \ref{fig:halpha_Hbeta} shows a Voronoi diagram illustrating the highly variable extinction $E(B$--$V)$ across the nebula. This map is for illustration purposes. To simplify the creation of this diagram from the MUSE data cubes, we have integrated the H$\alpha$ emission line between the two strong [\ion{N}{II}] lines  on either side and thus underestimate the H$\alpha$ emission. H$\beta$ was integrated over the same velocity interval. The extinction is highest close to MWC~137 and along the rim of the nebula and it decreases outside the nebula, clearly tracing the dust distribution, which is supported by 2MASS images of this region \citep{2006AJ....131.1163S}. 

\subsection{The extinction to MWC~137 via the spectral energy distributions of cluster stars}
\label{spt}

Figure \ref{fig:spectra} shows the spectra of the 39 brightest stars in the cluster, which have a sufficient signal to noise ratio for stellar classification. We determine a systemic cluster velocity of $50\pm20$~km~s$^{-1}$ based on the radial velocities of the Paschen lines (hot stars) and \ion{Ca}{II} and \ion{Ca}{I} lines (cool stars). The spectra have only few spectral lines and there are residuals from the nebula emission for the broad hydrogen lines. This results in less accurate radial velocity measurements than expected for the spectral resolution of MUSE. 

We determined the stellar spectral types with the help of the ESO UVES Paranal Observatory Project (POP) library  of high-resolution spectra ($R\sim80\,000$, \citealt{2003Msngr.114...10B}) and the Indo-US Library of low-resolution spectra ($R\sim5\,000$, \citealt{2004ApJS..152..251V}). The spectral resolving power of a range of template spectra, covering late-O to mid-G spectral subclasses, was decreased to $R \sim 3\,000$ to match our spectra. Equivalent widths of lines used for the spectral classification (see below) were measured in these template spectra and complemented by line equivalent widths listed in \citet{1987clst.book.....J} and \citet{1982A&AS...50..199D}.
MUSE does not cover the critical wavelength regions bluewards of 4750~\AA\  commonly used for stellar classification of hot stars \citep{1990PASP..102..379W}. The spectral range and resolving power do also not permit the classification of G-type star subclasses. Our spectral classification is thus only accurate to within a few spectral subclasses (Table \ref{table:spectraltype}). A review with spectra at bluer wavelengths is needed.

For the hot stars (spectral types B and A), we based our classification mostly on H$\alpha$ and H$\beta$ equivalent widths. The strength of \ion{O}{I} 7774 was used to confirm the spectral subclass.
The cluster does not contain any O-type stars. 
Two objects show weak \ion{He}{I} absorption and we determined their spectral type to early B.
For the cooler stars (spectral types F and later), we based our classification on the equivalent widths of \ion{Ca}{II} 8542 and \ion{Na}{I} 5889.  The second to last column in Table \ref{table:spectraltype} lists our spectral classifications.

In addition to MWC~137, we found two stars with potential H$\alpha$ emission (stars \#4 and \#16 in Table \ref{table:spectraltype}). H$\alpha$ and \ion{O}{I} 8446 emission remain after the nebular subtraction, while other nebular lines are reasonable well removed. This may indicate circumstellar material. The hydrogen Paschen lines are very broad in absorption and suggest a high value of $log~g$, indicating a luminosity class V and main sequence evolutionary phase. These two stars may be Be stars. However, the nebula emission varies strongly on small spatial scales and bad nebular subtraction may leave remnants that resemble narrow H$\alpha$ emission. Follow-up observations with higher spectral resolution are needed to determine whether these stars are indeed Be stars.

The right panel of Figure \ref{fig:fov} indicates the location of the cluster stars on sky. It demonstrates that the higher mass stars are located preferentially towards the cluster center, while the lower mass stars are found more towards the outskirts. Even though this qualitative result is somewhat biased by the higher extinction towards the center of the cluster, it indicates mass segregation commonly found in young clusters \citep{1993prpl.conf..429Z,1998MNRAS.295..691B}.

We compared the flux-calibrated MUSE spectra with the spectral energy distributions of UVES POP and Indo-US spectra and determined the extinction to each individual object. Using this method, we estimated on average a slightly lower extinction to the inner cluster region than by the Balmer emission lines decrement method described in Section \ref{balmer}. We found that $E(B$--$V) \approx 1.2 \pm 0.2$~mag is a reasonable extinction estimate throughout most of the inner 1\arcmin\ region of the cluster (Table \ref{table:spectraltype} and Figure \ref{fig:extinction_SpT}). This results in  $A_{V} \approx 3.7 \pm 0.6$~mag (adopting an extinction law with $R=3.1$) and $E(V$--$I) \approx 1.0 \pm 0.2$~mag. This result is similar to \citet{1998MNRAS.298..185E}, who determined $E(B$--$V)=1.216$~mag. Towards the outer parts of the cluster we find that the extinction drops to $E(B$--$V) \approx 0.4$--$0.8$~mag.\footnote{\citet{1996MNRAS.280..720V} had already measured different reddening corrections for the inner and the outer cluster regions.} In the center of the cluster, close to MWC~137, the extinction is higher. We have to adopt an extinction of $E(B$--$V) \sim 1.8$~mag ($A_{V} \sim 5.6$~mag) to match the spectrum of MWC~137 to a B0 \citep{2014MNRAS.443..947L} template spectrum. For comparison, using the Balmer decrement method described in Section \ref{balmer}, we found $E(B$--$V) \sim 1.5$~mag in the vicinity of MWC~137.

\begin{figure}
\centering 
\resizebox{\hsize}{!}{\includegraphics{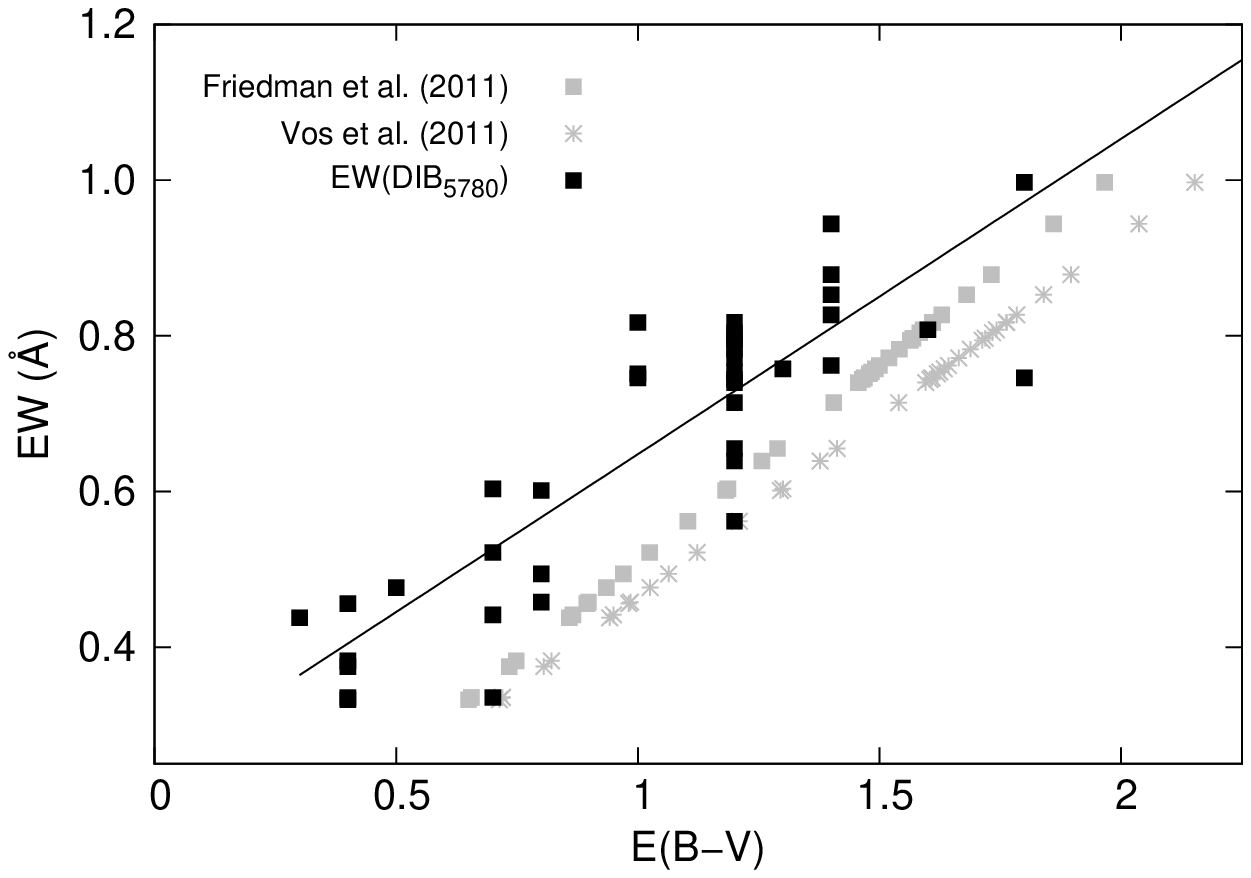}}
    \caption{The equivalent widths of the narrow component of the DIB at 5780~\AA\ correlates linearly with our $E(B$--$V)$ estimates determined in Section \ref{spt} (black squares). The $E(B$--$V)$ values calculated from our equivalent widths using the relations in \citet{2011ApJ...727...33F} and \citet{2011A&A...533A.129V} are also shown (gray points and gray stars). The offset of $\sim$0.25~mag between the calculated and our values is probably due to different equivalent measurement procedures. Our procedure includes some of the broad component of the DIB and we thus overestimate the equivalent widths.}
     \label{fig:DIB}
\end{figure}

The strengths of DIBs are closely correlated with extinction (e.g., \citealt{1993ApJ...407..142H,1994A&AS..106...39J,2007A&A...465..993G,2012A&A...544A.136R,2014A&A...569A.117C}). We measured equivalent widths of 0.3--1.0~\AA\ for the narrow component of the DIB at 5780~\AA\ in the stellar spectra towards the cluster SH~2-266. The equivalent widths show a linear correlation with the extinction estimates determined in this section (Figure \ref{fig:DIB} and Table \ref{table:spectraltype}), which confirms the wide range in $E(B$--$V)$. In addition, Figure \ref{fig:DIB} displays the $E(B$--$V)$ calculated using the relations in  \citet{2011ApJ...727...33F} and \citet{2011A&A...533A.129V}, which confirm our relative estimates. However, there is an offset of about 0.25~mag in the calculated $E(B$--$V)$ values with respect to the values determined in this section. This offset is probably due to different equivalent width measurement procedures. The DIB at 5780~\AA\ has a broad component, which, if included, results in larger equivalent widths by a factor of up to 1.5. Since we used only a rudimentary procedure, our measurements still include part of the broad component and we thus overestimate the equivalent width. Also, the value of $R_{V}$ depends on the environment along the line of sight and may be as high as 5 \citep{1989ApJ...345..245C}. In an extreme case, we would underestimate the extinction by up to 2~mag and the distance to the cluster would be several kpc further away from us.

\subsection{The distance, mass, and age of SH~2-266}
\label{distance}

\begin{figure}
\centering 
\resizebox{\hsize}{!}{\includegraphics{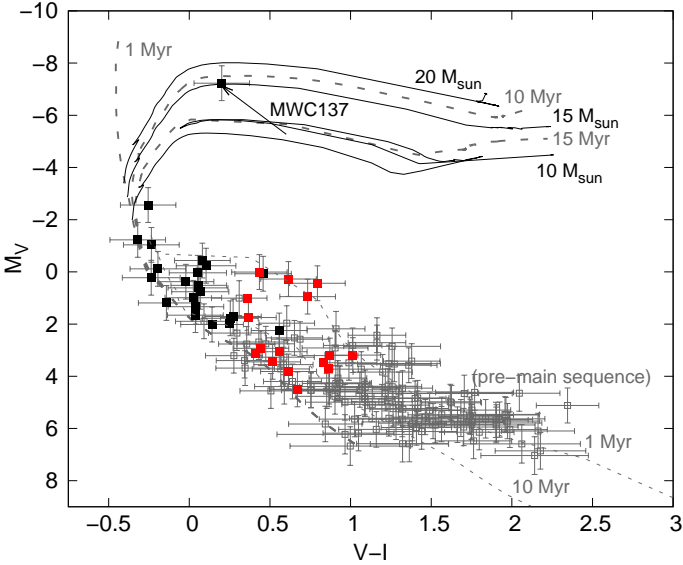}}
    \caption{$M_{\textnormal\scriptsize{{V}}}$ versus $V$--$I$ color-magnitude diagram of the brightest SH~2-266 cluster stars derived from the MUSE data. Black squares are B- and A-type stars, red squares are cooler F- and G-type stars, corrected for their individual extinction. The arrow at MWC~137 indicates the positions in the color-magnitude diagram from the average inner cluster estimate of $E(B$--$V)=1.2$~mag to the estimate for MWC~137 of $E(B$--$V) \sim 1.8$~mag. The open gray squares are unclassified stars corrected for a standard extinction of $E(B$--$V)=1.2$~mag. Theoretical stellar evolution isochrones for 1~Myr, 10~Myr, and 15~Myr (gray dashed curves) and the evolutionary paths for 10~$M_{\odot}$, 15~$M_{\odot}$, and 20~$M_{\odot}$ stars from the Geneva stellar models are shown \citep{2012A&A...537A.146E,2013MNRAS.433.1114Y}. 1~Myr and 10~Myr pre-main sequence theoretical isochrones for star masses of 0.1--7~$M_{\odot}$ from \citet{2000A&A...358..593S} are also shown (gray dotted curve).}
     \label{fig:v_vi}
\end{figure}

\begin{figure}
\centering 
\resizebox{\hsize}{!}{\includegraphics{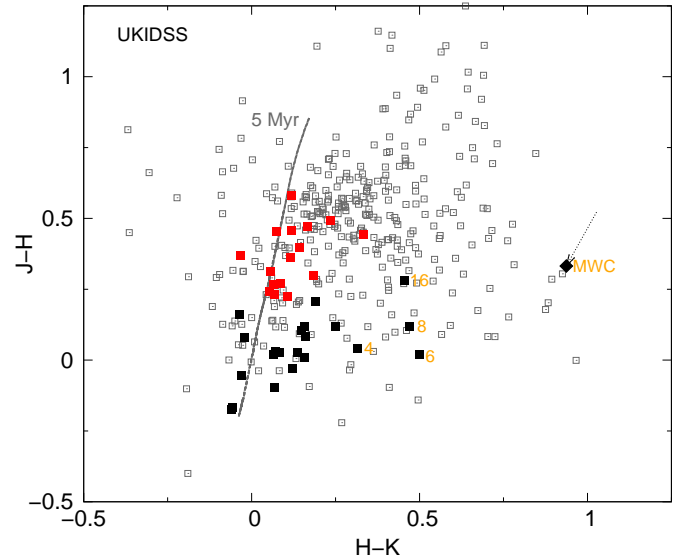}}
    \caption{$J$--$H$ versus $H$--$K$ color-color diagram from the UKIDSSDR8PLUS database. Black squares are B- and A-type stars, red squares are cooler F- and G-type stars. The open gray squares are unclassified stars. The theoretical stellar evolution isochrone for 5~Myr from the Geneva stellar models is shown \citep{2012A&A...537A.146E,2013MNRAS.433.1114Y}. MWC~137, \#10, \#23, \#25, and \#28 are saturated in the UKIDSS images and their $JHK$ values were obtained from the 2MASS point source catalogue (\citealt{2003yCat.2246....0C}; the 2MASS catalogue does not contain the star \#10, which is thus not displayed here). The data were dereddened adopting the $E(B$--$V)$ values determined in Section \ref{spt} and the relations $E(J$--$H)/E(H$--$K)=2.09$~mag \citep{1980MNRAS.192..359J} and $E(J$--$H)/E(B$--$V)=0.33$~mag \citep{1970ApJ...162..217L}. For MWC~137 the dereddening vector is shown from $E(B$--$V)=1.2$~mag to $1.8$~mag. For the unclassified stars a default value of $E(B$--$V)=1.2$~mag was assumed.}
     \label{fig:ukidss}
\end{figure}

Figure \ref{fig:v_vi} shows the absolute visual magnitude $M_{V}$ versus $V$--$I$ of the SH~2-266 cluster stars derived from the MUSE data. Magnitudes and colors were corrected for the individual extinctions determined in Section \ref{spt}. The distance modulus was determined to $\mu = 13.6 \pm 0.8$~mag by minimizing the difference in $M_V$ for all stars to literature values for the respective spectral types \citep{2006MNRAS.371..185W,2007MNRAS.374.1549W}.
The errors in both color and absolute magnitude are large, because of the high uncertainties in the interstellar medium extinction and the uncertainties in the spectral classification (see Sections \ref{balmer} and \ref{spt}). 

A distance modulus of $\mu = 13.6$~mag  results in a good fit of the main sequence cluster stars with Geneva stellar evolution models \citep{2012A&A...537A.146E,2013MNRAS.433.1114Y}. This corresponds to a distance of about $5.2\pm1.4$~kpc. MWC~137 is located in the Galactic plane, almost opposite from the Galactic center ($l=195.6502\degree$, $b=-0.1046\degree$). At a distance of about 2.5~kpc in this direction lies the Perseus arm and at about 6~kpc the Cygnus arm \citep{2009PASP..121..213C,1993ApJ...411..674T}. SH~2-266 may thus be associated with the Cygnus arm.

Evolutionary tracks of 10--15~$M_{\odot}$ stars fit well the location of MWC~137 in the color-magnitude diagram (Figure \ref{fig:v_vi}). The two early B-type stars have masses of 6--8~$M_{\odot}$  \citep{2007A&A...471..625T,2007A&A...466..277H}. If we adopt a total mass of about $25~M_{\odot}$ for the three most massive stars, a Salpeter initial mass function \citep{1955ApJ...121..161S} results in a total mass estimate of stars with masses $\geq1~M_{\odot}$ in this cluster of about $100~M_{\odot}$. We obtain the same result by adding up the masses of the stars in Table \ref{table:spectraltype}. The mass function thus appears to be consistent with the initial mass function.

From the color-magnitude diagram of the brightest cluster members (Figure \ref{fig:v_vi}) we determined an upper limit to the age of the cluster of about 30~Myr. This conservative estimate is based on the assumption that MWC~137 is not following the evolutionary track of a single star. In this case, the main sequence cut-off of this cluster is at about $8~M_{\odot}$ (stars with spectral types B2--4 and later lie along the main sequence). The estimation of a lower age limit is difficult. In the case MWC~137 is following the post-main sequence evolution of a single star, its mass can be estimated to about 10--15$~M_{\odot}$ and its age to about 10--15~Myr. We did not find any evidence of an object in this cluster that still possesses its protoplanetary disk down to a mass of about 1~$M_{\odot}$. Therefore, the cluster age is likely above 1--3~Myr \citep{2015A&A...576A..52R}.

Figure \ref{fig:ukidss} shows a near-infrared color-color diagram from the UKIDSS survey. The near-infrared colors confirm that the cluster is at least about 3~Myr old. A few high-mass stars are observed to have excess near-infrared emission (compare to \citealt{1983A&A...128...84K}) possibly due to hot dust and in the case of MWC~137 a disk-like structure may also contribute. Near-infrared emission is present for the stars \#4 and \#16, confirming the presence of circumstellar material, which the H$\alpha$ emission in their spectra suggested (Section \ref{spt}). For the two other stars with near-infrared excess, stars \#6 and \#8, there is no evidence of optical line emission.

Figure \ref{fig:v_vi} indicates that a few F- and G-type stars may not have reached the main sequence yet, but have (pre-main sequence) ages between 1~Myr and 10~Myr based on the tracks by \citet{2000A&A...358..593S}. These stars (\#24,25,30,31 at $M_{V}\sim 0.5$~mag) are located along the north-east rim of the nebula, indicating a possible age spread. A shock front originating from MWC~137 could have caused a second episode of star-formation in this cluster. However, the stars \#24 and \#25 have very bright $V$-band magnitudes, a low reddening, and a slightly different radial velocity, and are thus likely not cluster members. The near-infrared data show no evidence for disks around them.  Note that most F- and G-type stars show less reddening than the B- and A-type stars, consistent with the strengths of their DIBs. This implies that these stars are either foreground stars, which explains the higher $M_V$ magnitudes compared to literature values for their spectral types (Table \ref{table:spectraltype}), or are found preferably at the front side of the cluster.

To summarize the different possibilities related to the age of the cluster and its very probable member MWC~137:
 
\begin{enumerate}
\item MWC~137 is following the evolution of a single massive star. Its position in the Hertzsprung-Russell (HR) diagram is best fit with 10--15~Myr isochrones, which have a main sequence turn-off around 10--15$~M_{\odot}$. This is in agreement with the near-infrared colors of the cluster stars.
 
\item MWC~137 is not following the evolution of a single star. The HR diagram is best fit with a 30~Myr isochrone, where the main sequence cut-off is around $8~M_{\odot}$. This is also in agreement with the near-infrared colors of the cluster stars.
 
\item MWC~137 is not following the evolution of a single star. Based on the fact that no stars with protoplanetary disks are found in this cluster and the near-infrared colors, the cluster is older than 3--5~Myrs.

\item The cluster may have had several star-formation episodes. In this case a single age determination is meaningless.

\end{enumerate}

\begin{figure}
\centering 
\resizebox{\hsize}{!}{\includegraphics{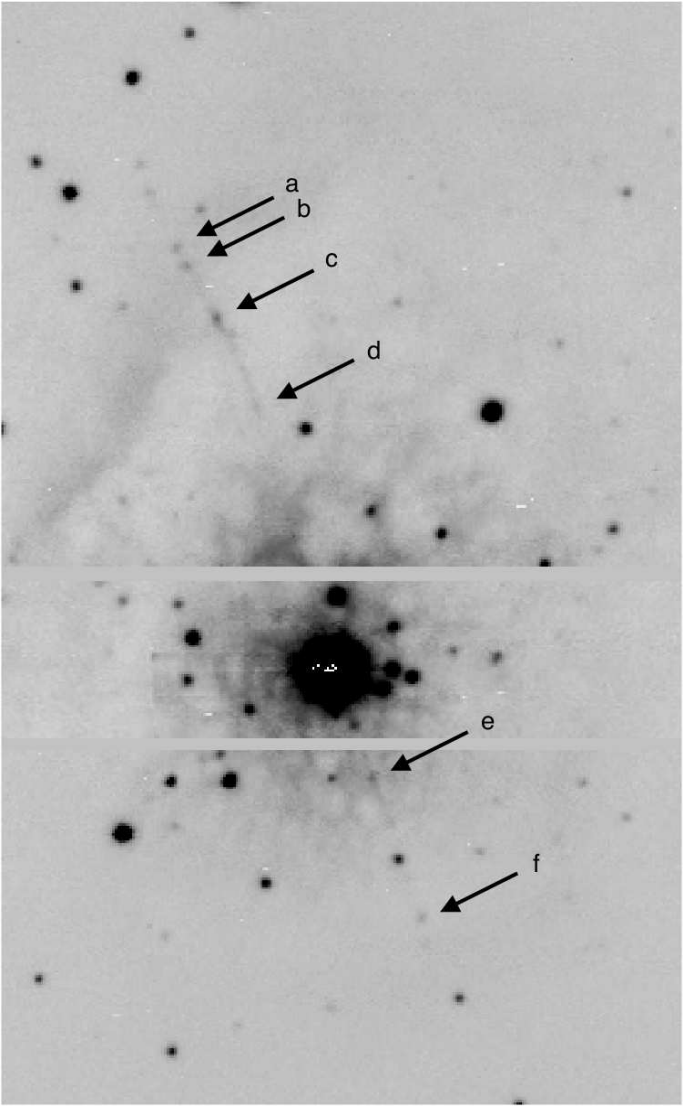}}
    \caption{Composed image showing the jet in [\ion{N}{II}] 6583. The upper and lower panels are velocity-binned images from 600~s exposures ($-$405~km~s$^{-1}$ to $-$234~km~s$^{-1}$ for the upper and 336~km~s$^{-1}$ to 507~km~s$^{-1}$ for the lower panel). The middle panel is one channel map (at $-$376~km~s$^{-1}$) of a 300~s exposure. Knots along the jet trace are labelled `a' to `f'. Other knot-like features coincident with the jet trace have stellar spectra.}
     \label{fig:jetI}
\end{figure}

\subsection{The jet}
\label{jet}

\begin{figure}
\centering 
\resizebox{\hsize}{!}{\includegraphics{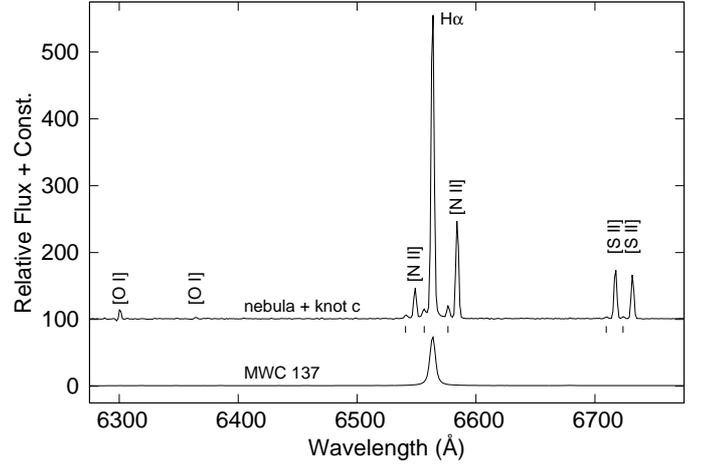}}
    \caption{Spectrum of knot `c' and the nebula. The strong [\ion{N}{II}], [\ion{S}{II}], and [\ion{O}{I}] emission lines are from the extended nebula. The weaker emission at about $-320$~km~s$^{-1}$ with respect to these nebular lines are the emission lines originating from the knot (indicated by small marks).  \label{fig:knot}}
\end{figure}

We discovered a collimated outflow in the MUSE data that appears to originate from MWC~137 at a position angle of 18--20\degree\ (Figure \ref{fig:jetI}). We detect the jet in H$\alpha$, [\ion{N}{II}] 6548,6583, and [\ion{S}{II}] 6716,6731, but not in [\ion{O}{I}] 5577,6300. 
The extension of the jet as traced by the [\ion{N}{II}]  6583 line is about 66\arcsec, which corresponds to a length of 1.7~pc projected on the plane of sky, if we adopt a distance of 5.2~kpc. The width of the jet is with 3 pixels not resolved, but corresponds to the seeing of about 0.6\arcsec.  The jet extends almost across the entire cluster (Figures \ref{fig:fov} and \ref{fig:jetI}). The jet outflow, especially the northern outflow, shows deviations from a linear trace and at least six knots can be identified (labelled `a' to `f' in Figure \ref{fig:jetI}). This may hint at a velocity variability and a precession of the jet axis \citep{2001A&A...367..959R}.

The velocity structure within the jet is resolved over about five MUSE channels. The jet material expands with projected velocities between $-148$~km~s${^{-1}}$ to $-433$~km~s${^{-1}}$ towards the north and between $+250$~km~s${^{-1}}$ to $+478$~km~s${^{-1}}$ towards the south.  A slowing down of the jet with increasing  distance to MWC~137 is observed. 
Given the large spatial extent and the large velocities, we assume an inclination of the jet to the plane of the sky close to 45\degree. Then the de-projected jet length is 93\arcsec\ (2.3~pc, assuming a distance of 5.2~kpc) and the de-projected speeds along the jet are up to 650~km~s$^{-1}$.  
 Note that we do not have any firm constraint on the inclination. For a larger inclinations, the velocities within the jet may be well above 1\,000~km~s$^{-1}$.  Adopting a speed of material in the jet of 650~km~s$^{-1}$ and a travel distance of 1.2~pc, the life time of the outermost part of the jet is about 1800~yr. As comparison, we estimate the age of the large nebula to $\geq$20\,000~yr, based on an upper limit of its expansion speed of 60~km~s$^{-1}$ (limited by the spectral resolving power of MUSE) and its extension of about 50\arcsec\ from MWC~137 at a distance of 5.2~kpc. Thus, the jet did not originate in the same mass-loss episode that caused the nebula, but is much younger.

Figure \ref{fig:knot} shows a spectrum of knot `c' in the H$\alpha$ region. The strong nebular [\ion{N}{II}], [\ion{S}{II}], and [\ion{O}{I}] dominate this part of the spectrum. Other strong nebular lines are H$\beta$ and Paschen lines, [\ion{O}{I}] 5577, and [\ion{S}{III}] 9069, weaker emission can be seen from [\ion{N}{I}] 5200, \ion{He}{I} 5876, 6678, [\ion{Ar}{III}] 7136, and [\ion{O}{II}] 7319,7330,7331. The figure indicates with marks the weaker emission from knot `c' at about $-320$~km~s$^{-1}$ with respect to the nebular lines. A comparison spectrum of MWC~137 shows that the large nebula is not a reflection nebula, but is ionized by the bright central source.

The detection of the jet in [\ion{N}{II}], but not in [\ion{O}{I}] implies temperatures larger than $10\,000$~K \citep{1994ApJS...93..485H}. 
The electron density can be estimated by the ratio of the [\ion{S}{II}] lines at $\lambda\lambda$6716,6731. 
We found a flux ratio [\ion{S}{II}] 6716/6731 of  about $1.1$ for knot `a', $1.0$ for knot `b', and $0.8$ for the knot `c'. This implies that the electron density is decreasing with increasing distance to the star (as expected) and that the absolute value is $n_e \sim 10^3$~cm$^{-1}$ \citep{2006agna.book.....O}. A rough estimate of the mass-loss rate within the jet results in $\dot{M} = 2 \times \pi r^2 \rho v \sim 10^{-8}~M_{\odot}$~yr$^{-1}$, assuming a knot radius of $r = 0.7$\arcsec, an ionization fraction of 20\% \citep{1999A&A...342..717B}, and a velocity of $v = 650$~km~s$^{-1}$. If the jet originated in an accretion episode and we assume that 10\% of the accreted mass is ejected in the jet, then we obtain an upper limit for the  accretion episode of about $10^{-7}~M_{\odot}$~yr$^{-1}$.

\section{Discussion}
\label{discussion}

The central position of MWC~137 in the cluster SH~2-266, within the large nebula, and also with respect to the jet, suggests strongly that MWC~137 belongs not only to this cluster, but is also at the origin of the nebula and the jet. 

Collimated outflows occur in most astrophysical systems in which accretion, rotation, and magnetic fields interact.  
Jet physics are still highly uncertain, especially the jet generation mechanism. However, it is widely accepted that a fast, highly collimated jet requires an accretion disk. The stronger the accretion, the more powerful the jet. \citet{2008ApJ...689.1112C} showed that 1--7\% of the accreted mass is ejected in outflows for T~Tauri jets. Similar values are found for HAeBe stars (e.g., \citealt{2014A&A...563A..87E}).

The outflow symmetries provide clues about the dynamical environment of the engine. The jet symmetry in MWC~137 deviates from a linear trace (Figure \ref{fig:jetI}). This indicates that the outflow axis changed over time, maybe due to a precession induced by a companion, motion of surrounding gas, or moving of the outflow source itself.
We observe several knots along the jet (four in the northern outflow and two in the southern outflow). These knots may form where faster ejecta run into slower material, but may also indicate variations in the mass ejection velocity or density. Variations in the low-density outbursts seems more likely for MWC~137's jet, because the knot spectra are dominated by the same low-excitation lines as the nebula and we do not detect significant line ratio variations of H$\alpha$/[\ion{S}{II}] and H$\alpha$/[\ion{N}{II}] between the knot and the nebula spectra. Photoionization thus dominates the nebula and the jet emission spectra.  Apart for knots `d' and `f', we do not find any point symmetry in the location of the knots with respect to MWC~137, which would indicate a common event. We estimate a density of $\sim$10$^3$~cm$^{-3}$. The inner jet components are denser, hotter, and faster.  Average jets have densities of $10^2$ to $10^5$~cm$^{-1}$ \citep{1999A&A...342..717B}. Based on the spatial extent and outflow velocities, we estimate a minimum life of the jet of at least 1800~yrs (compared to an age of $\geq 20\,000$~yr for the nebula). 

\citet{1999MNRAS.305..166O} used spectropolarimetry across H$\alpha$ of MWC~137 and derived an intrinsic polarization angle of 30\degree. 
This polarization supports the presence of a dense H$\alpha$-emitting aspherical structure such as a disk around the star. Typically, the intrinsic polarization angle is perpendicular to the circumstellar disk structures \citep{2002MNRAS.337..356V,2005A&A...430..213V}, and therefore the disk-like structure for MWC~137 has a position angle on sky of 120\degree, i.e., perpendicular to the jet. This supports the idea that the jet is associated with a disk around MWC 137. 
The absence of hydrogen P~Cygni profiles in the spectrum of MWC~137 suggests that the disk-like structure has some inclination to the plane of the sky \citep{1985A&A...143..421Z}.

Infrared data suggest that the disk of MWC~137 is probably relatively compact. Although, the near-infrared and mid-infrared colors demonstrate the presence of hot and warm dust (see Figure \ref{fig:ukidss} and \citealt{2014AdAst2014E..10D}), MWC~137 does not show strong evidence for a disk-like environment at long wavelengths.
\citet{2011ApJ...727...26S} saw no enhancement in cold dust emission towards MWC~137 in their $850~\mu m$ SCUBA maps. These authors concluded that MWC~137 has photoionized most of its disk and is more evolved than a HAeBe star. 
\citet{2004Ap&SS.292..465F} found also no compelling evidence for a disk. They determined an upper limit of $0.007~M_{\odot}$ to the mass of a circumstellar disk based on its spectral energy distribution at mm and cm wavelength obtained with the IRAM and the VLA interferometers. 

The origin and nature of the jet in MWC~137 is not clear. While we have found that MWC~137 is a sgB[e], its {\em evolutionary history} is unknown. In particular we do not know the initial and current surface rotational speed, mass-loss history, and binarity status. These factors significantly affect the evolution of the star to a point of invalidating the evolutionary tracks shown in Figure \ref{fig:v_vi}. Nevertheless, we discuss below several possibilities.

\begin{enumerate}

\item {\it Circumstellar material due to a close binary star.} The majority of massive stars are born in multiple systems \citep{2008MNRAS.386..447S,2012MNRAS.424.1925C,2013A&A...550A.107S,2014ApJS..213...34K}. \citet{2015arXiv150507121D} found a present-day binary fraction of $58\pm11$\% for B-type stars in the 30 Doradus region of the Large Magellanic Cloud. \citet{2012MNRAS.424.1925C} found that more than 80\% of stars with masses above $16~M_{\odot}$ in the Milky Way form close binary systems. The presence of a companion significantly influences the evolution of massive stars \citep{1992Natur.359..305P,2008IAUS..250..167L,2011ASPC..440..217E} and the widespread importance of stellar mass transfer in binary systems has only been realized recently.
In the case of MWC~137, circumstellar material may accumulate in the Roche lobe of a putative (smaller) companion to form a disk-like structure, from which a jet could be launched  \citep{1996A&A...312..879H,2007A&A...463..233A,2015arXiv150205541A}. In such disks, one does not expect any dust, consistent with the lack of dust emission found for MWC~137. Another possibility is that the system is similar to the B[e] star CI CAM, which experienced an outburst and has likely either a neutron star, black hole, or white dwarf companion \citep{1998A&A...339L..69F,2006ASPC..355..305B}. 
The disk-like feature may also be bound material from a previous mass transfer of a hotter companion, which may now be an undetected white dwarf. To have a white dwarf companion to a 10--15~$M_{\odot}$ star, the mass exchange would have involved at least three solar masses.

\item {\it Accretion disk from a stellar merger.}  Stars that appear to be single may have evolved from binary systems \citep{2014ApJ...782....7D}. About 24\% of all O-type stars may be merged binaries \citep{2012Sci...337..444S}. This has important consequences on their inferred evolutionary history \citep{2013ApJ...764..166D}. MWC~137 may possibly be the product of a stellar merger of two B-type stars. The large circumstellar nebula may have originated in this merger event around $20\,000$~yr ago. In addition, an accretion disk may have formed around the merged massive object, which is still fueling the jet. 
This scenario could explain the unprocessed nebula material, the disk-like structure and the jet. 
However, the mass function of the cluster appears to be consistent with the initial mass function without having to invoke a merger scenario. 

\item {\it Protoplanetary disk.} When stars form, an accretion disk develops after an initial radial infall of material. Disk life times are mass dependent, i.e., they are shorter for higher mass stars \citep{2004Ap&SS.292..465F,2015A&A...576A..52R}. Magnetic fields couple the disk to the star and an accretion system drives a strong outflow that can lead to a jet. A few HAeBe stars are associated with jets (e.g., \citealt{2014A&A...563A..87E}). The estimated protostellar jet phase is about 10$^5$~yr.
However, MWC~137 is the most massive star in the cluster with a mass of about 10--15~$~M_{\odot}$, while other cluster members have masses $\lesssim 8~M_{\odot}$. 
If MWC~137's disk is the evolved remnant from the star's protoplanetary phase then some physical process, e.g., a strong magnetic field, must have conserved it, because we do not observe disks around any of the later type stars. 
If MWC~137 is a pre-main sequence star, it could belong to a younger generation of stars in this cluster, but this hypothesis is highly speculative.

\item{\it Planetary nebula phase.} In the planetary nebula stage, jets are generally much less collimated than is the case for MWC~137 and show smaller outflow velocities \citep{1987AJ.....94.1641B}. However, some planetary nebula show relatively narrow jets, such as Fleming~1 \citep{1993A&A...267..194L,1993ApJ...415L.135L}, where a binary shapes the point-symmetric jet \citep{2012Sci...338..773B}. 

\item {\it Jet originates from star `4'.} Connecting the end points of the observed collimated jet with a linear curve MWC~137 is the most likely choice for its origin (Figure \ref{fig:fov}) and both outflows appear to trace directly back to MWC~137. However, we find H$\alpha$ emission and thus evidence for circumstellar material in the nearby star \#4. 

\item {\it Jet arises from a less massive, distant, unresolved companion, which does not interact with MWC~137.} The closest star to MWC~137 that we can resolve in our MUSE data is at about 15\,000~au (assuming a distance of 5.2~kpc). Thus, there is the possibility that a nearby star or a wide companion star with no interaction with MWC~137 may be the origin of the jet. The properties of such a companion cannot be observationally constrained from our data, but it has to be cool enough to not excite much line emission in the disk. 

\end{enumerate}

\section{Conclusion}
\label{conclusion}

We have presented a study using VLT MUSE IFU data of the host star cluster of the enigmatic early B[e]-type star MWC~137. Our analysis of the optical color-magnitude diagram and the IFU spectra of the cluster stars has allowed a distance estimate of $5.2\pm1.4$~kpc.
The data establish that  MWC~137 is the most massive member of the cluster with a spectroscopically identified main sequence populated by A and B-type stars. Based on isochronal analysis we can conclude that MWC~137 is evolved off the main sequence. Its high luminosity of $log~(L/L_{\odot}) \sim 6.0$ and its {\it IRAS} colors \citep{1988iras....7.....H} show that it is not a member of the FS~CMa class \citep{2007ApJ...667..497M}, but a sgB[e] star. 

Despite a firm characterization of MWC~137 as an evolved object, we also present the surprising discovery of a jet centered on MWC~137. 
The jet has projected velocities of up to $\pm450$~km~s$^{-1}$ and a projected length of 66\arcsec. Several knots are seen along the jet, which originate either from the
interaction of the fast jet with the slower surrounding material or
from velocity variations within the outflow. Jets of this morphology
are usually associated with young star accretion from a primordial
disk. An H$\alpha$ spectro-polarimetric signal supports the presence
of a gaseous disk associated with MWC~137, yet this disk contains
little to no cold dust. It is therefore debatable whether the
jet-engine is a primordial disk or a disk generated by an interactive
phenomenon. This uncertainty is nourished by the fact that the cluster
members with spectral types as late as approximately G5V display
little to no evidence for primordial disks. We find that the outermost
observed knots of the jet are about 1800~yr old, whereas the larger
surrounding nebula is limited in age to $\geq$20\,000~yr. As an
alternative origin for the jet, we speculate that during a high
mass-loss episode that lead to the formation of the nebula (e.g., an
eruption or a merger event), an accretion disk formed that fuels the
jet. Another possibility is that MWC~137 has either a neutron star,
black hole, or white dwarf companion, similar to the B[e] star CI~CAM. Table \ref{table:1}
summarizes our findings on the cluster, nebular, stellar, and jet
parameters.


The MUSE IFU data of the cluster SH~2-266 have provided important new insights into the evolution of its most massive member. The data allowed us to strongly constrain the distance to the cluster and unambiguously demonstrate that MWC~137 is a post-main sequence object based on its position in the HR diagram. Yet, the enigmatic nature of this source has only increased by the discovery of a fast jet emanating from it. Future observations with instruments probing smaller spatial scales and different wavelengths will let us analyze the jet and the disk-like structure in more detail. Deeper observations will be required to detect the main sequence turn-on and determine the age of the cluster by means of pre-main sequence evolutionary tracks.

\begin{table}
\caption{Derived cluster, nebula, stellar, and jet parameters.}              
\label{table:1}      
\centering                                      
\begin{tabular}{l c}          
\hline\hline 
\multicolumn{2}{c}{SH~2-266}  \\   
$E(B$--$V)$ &   $1.2\pm0.2$~mag \\[2.5pt]   
$A_V$ & $3.7\pm0.6$~mag \\[2.5pt]     
$d^a$ & $5.2\pm1.4$~kpc \\ 
age &  	$3$~Myr~$\lesssim t \lesssim15$~Myr$^b$ \\
mass & $>100~M_{\odot}$\\
nebular age & $>20\,000$~yr\\
$v_{systemic}$ & $50\pm20$~km~s$^{-1}$ \\\hline 
\multicolumn{2}{c}{MWC~137}  \\                        
$E(B$--$V)$ &   $\sim$1.8~mag \\[2.5pt]   
$A_V$ & 5.6~mag \\[2.5pt]      
M$_V$ & $-7.2$~mag \\
$log~(L/L_{\odot})^c$ & 6.0 \\
mass & 10--15~$M_{\odot}$\\ 
\hline 
\multicolumn{2}{c}{Jet}  \\                        
projected length  &	66\arcsec\ (1.7~pc) \\
de-projected length$^d$  &	93\arcsec\ (2.3~pc) \\
velocity & 	$-433$~km~s$^{-1}$ to $+478$~km~s$^{-1}$\\
de-projected velocity$^d$ & 	up to $650$~km~s$^{-1}$\\
age & $>1800$~yr \\
electron density & $\sim10^3$~cm$^{-3}$\\
mass-loss rate & $\sim 10^{-8}~M_{\odot}$~yr$^{-1}$ \\
\hline
\multicolumn{2}{l}{$^a$In the future Gaia will provide a very accurate parallax}  \\ 
\multicolumn{2}{l}{\phantom{$^a$}for this object. Hipparcos found a parallax of about}  \\                                            \multicolumn{2}{l}{\phantom{$^a$}1~kpc \citep{2007A&A...474..653V}.}  \\                                           
\multicolumn{2}{l}{$^b$Assuming a single star evolution for MWC~137.}  \\                                            \multicolumn{2}{l}{$^c$Using a bolometric correction of $BC=-3.03$ (for a B0}  \\                                             \multicolumn{2}{l}{\phantom{$^c$}star; \citealt{2013ApJS..208....9P}).}  \\
 \multicolumn{2}{l}{$^d$Assuming an inclination angle of 45\degree.}  \\                                             
\end{tabular}
\end{table}

\begin{acknowledgements} This research has made use of the VizieR catalogue access tool, CDS, Strasbourg, France. We have also used UKIDSS data from the UKIDSSDR8plus database. The UKIDSS project is defined in \citet{2007MNRAS.379.1599L}. UKIDSS uses the UKIRT Wide Field Camera (WFCAM; \citealt{2007A&A...467..777C}) and a photometric system described in \citet{2006MNRAS.367..454H}. We thank R.M.\ Humphreys for her comments on an early version of the paper and N.L.J.\ Cox for his advice on DIBs.
\end{acknowledgements}

\bibliographystyle{aa}

\begin{landscape}
\begin{table}
\caption{Spectral classification for the brightest cluster members.}              
\label{table:spectraltype}      
\centering                                      
\begin{tabular}{c c c c c c c c c c c c c c c c l}          
\hline\hline                         
\# &  R.A. & Dec. & $V$ & $V$--$I$ & $M_V^a$ & EW & EW & EW &	EW &	EW &	EW	 &	EW & $E(B$--$V)$&$E(B$--$V)$ & SpT & Remark \\  
 &   &  &  &  &  & H$\alpha$ & ${H\beta}$ & CaII$_{8542}$ &	OI$_{7774}$ &	NaI$_{5889}$ &	DIB$_{5781}	$ &	DIB$_{5797}$	& 	${B.D.}^b$	&${SpT}^c$	&  &  \\  
 &   (h m s) & (\degree\ \arcmin\ \arcsec)  & (mag) & (mag) & (mag) & (\AA) & (\AA) & (\AA) &	(\AA) & (\AA) &	(\AA) & (\AA)  &	 (mag)	& (mag)	& &  \\[2.5pt]   
\hline 
1  &	6 18 45.52	&	15	16	52.25	&	11.95	& 1.64 &	$-$7.23	&	$-$611.71	&	$-$71.67	&	-	&	$-$2.29	&	0.27		&	0.75	&	0.37	& 1.5	& $\sim$1.8 &	sgB[e]$^d$ 	& MWC~137 \\
2 &	6 18 43.31	&	15 16 42.44	&	14.76	& 0.71	&	$-$2.56	&	6.57		&	6.72	&	-	&	0.33		&	0.68		& 0.80	&	0.24	& 1.5	&	1.2	&	B2--4V&	He I abs. \\
3 &	6 18 44.64	&	15 17 13.36	&	16.09	&	0.64&	$-$1.23	&	7.45		&	8.67		&	-	&	0.43		&	0.64	  & 0.82	&	0.26& 1.2	&	1.2 &	B3--5V	& He I abs. 	\\
4  &	6 18 45.50	&	15 16	58.29	&	16.91	& 0.85	&	$-$1.03	&	$-$9.38	&	11.71	&	-	&	0.70	&	0.81		& 0.76	&	0.25		& 1.5&  1.4	&	B8--9V& H$\alpha$ em.		\\
5 &	6 18 46.30	&	15 16 54.97	&	17.20	&	0.77&	$-$0.12&	9.71		&	8.71	&	-	&	0.04		&	0.77	 &	0.78	&	0.26	& 1.5	&	1.2		&	B6--8V&\\
6 &	6 18 45.24	&	15 16 50.75	&	17.35	& 1.01	&	0.03&	12.15	&	9.15	&	-	&	0.37		&	0.89 	& 0.77&	0.23	& 1.4	&	1.2		&	B7--8V&	\\
7 &		6 18 46.10	&	15 16 43.25	&	17.70	&	0.94	&	0.38&	10.68	&	10.00	&	-	&	0.40		&	0.81	& 0.74&	0.25		& 1.5	&	1.2	&	B8--9V	&	\\
8 &	6 18 45.19	&	15 16 52.34	&	17.70	& 1.22	&	$-$0.24&	7.38		&	8.69		&	-	&	0.25		&	0.74	&	 0.83	&	0.24	& 1.5	&	1.4	&	B7--8V	&\\
9 &	6 18 48.40	&	15 16 51.23	&	17.83	& 1.02	&	0.51	&	8.99		&	11.32	&	-	& 0.64	&	0.81		& 0.75	&	0.23	& 1.5	&	1.2	&	B8--9V&	\\
10 &	6 18 45.08	&	15 16 51.74	&	17.93	& 1.01	&	0.61&	12.94	&	11.56	&	-	&		0.49	&	0.64	& 0.74	&	0.25	& 1.4	&	1.2	&	B8--9V	&	\\	
11 &	6 18 42.52	&	15 16 54.46	&	18.03	& 0.84	&	1.33&	14.55	&	16.18	&	-	&		0.83		&	0.78		& 0.82	&	0.24		& 1.4	&	1.0&	A1--3V	& 	\\
12 &	6 18 47.50	&	15 17 16.17	&	18.12	& 1.36	&	-0.44&	8.89		&	9.56	&	-	&	0.57		&	0.86			& 0.81	&	0.22	& 1.6	&	1.6	&	B8--9V&	\\
13 &	6 18 45.19	&	15 16 55.79	&	18.16	& 0.88	&	0.22&	6.73		&	12.45	&	-	&	0.23		&	0.75	& 0.88	&	0.28& 1.5 &	1.4 	&	B8--9V&		\\	
14 &	6 18 42.10	&	15 16 56.77	&	18.36	& 0.84	&	1.66&	13.41	&	15.74	&	-	&	1.12		&	0.34		& 0.75	&	0.23		& 1.4	&	1.0	&	A3--4V &	\\
15 &	6 18 45.68	&	15 17 12.01	&	18.51	&	0.82&	1.19&	10.58	&	13.62	&	-	&	0.94		&	0.64			& 0.79	&	0.27	& 1.5	&	1.2&	B9--A0V	& \\
16 &	6 18 46.34	&	15 16 51.53	&	18.62	&	1.06	&	0.99&	2.86		&	13.58	&	-	&	0.49		&	0.87			&	0.76	&	0.20	& 1.4	& 1.3 &	B9--A1V& H$\alpha$ em.$^e$	\\
17 &	6 18 44.48	&	15 16 16.53	&	18.66	& 1.05	&	1.96&	14.30	&	15.18	&	-	&	0.39		&	0.64		& 0.75	&	0.25	& 1.5	&	1.0	&	A0--4V& 	\\
18 &	6 18 45.99	&	15 16 49.14	&	18.70	& 1.19	&	0.76&	13.22	&	15.18	&	-	&	0.31		&	0.99	& 0.94	&	0.20	& 1.4	&	1.4		&	A0--4V& 	\\
19 &	6 18 44.33	&	15 17 00.82	&	19.03	&	1.23&	1.71	&	11.64	&	9.95	&	-	&	0.35		&	0.84		& 0.66	&	0.22		& 1.3	&	1.2	&	A7--9V& 	\\
20 &	6 18 47.19	&	15 17 33.87	&	19.10	& 1.21	&	1.78&	15.78	&	16.36	&	-	&	1.01		&	0.64		& 0.64	&	0.26	& 1.6	&	1.2	&	A1--2V& 	\\
21 &	6 18 46.42	&	15 16 43.23	&	19.25	&	1.90	&	0.07	&	12.23	&	10.68	&	-	&	0.78		&	1.08		& 1.00	&	0.35	& 1.4	&	1.8	&	A1--2V&	\\
22	&	6 18 46.39	&	15 16 57.66	&	19.34	&	1.10	& 2.02	&		16.68	&	13.61	&	-	&	1.14		&		0.91	&	0.71	&	0.13		&  1.5	&	1.2	& A1--3V	&	\\
23	&	6 18 46.86	&	15 17 49.45	&	19.57	& 1.52	& 2.25	&		6.25	&	8.82	&	-	&	0.50		&		0.96		& 0.56	&	0.12		& 1.4	&	1.2	&	A9--F0V& 	\\
\hline
24	&	6 18 48.50	&	15 16 46.19	&	14.80	& 0.86	&	0.27	&	2.16	&	5.05	&	2.89	&	0.01		&	0.95	&	0.44&	0.11	& 1.5	&	0.3		&	F6--8V&	fg$^f$ \\
25	&	6 18 46.05	&	15 17 48.30	&	14.85	& 0.75	&	0.01	&	4.09	&	4.45	&	2.83	&	0.10		&	1.11		&	0.46	&	0.13& 1.2	&	0.4	&	F6--8V&	fg$^f$	\\
26	&	6 18 43.80	&	15 16 14.30	&	15.85	& 0.68	&	1.01&	5.87	&	6.39	&	2.07	&	0.35		&	0.79		&	0.33&	0.11	& 1.6	&	0.4	&	F1--3V&		\\
27	&	6 18 46.69	&	15 16 38.95	&	16.59	&	0.69&	1.75&	8.86	&	6.71	&	2.25	&	0.05		&	0.84		&	0.38&	0.11& 1.5	&	0.4	&	F1--3V&		\\
28	&	6 18 43.89	&	15 16 05.99	&	17.90	& 0.88	&	3.06&	2.18	&	4.32	&	2.98	&	0.14		&	1.38		&	0.37&	0.07	& 1.6	&	0.4	&	F6--8V&		\\
29	&	6 18 46.64	&	15 17 40.73	&	18.08	& 0.84	&	2.93&	4.88	&	5.93	&	2.16	&	0.30		&	1.01		&	0.48&	0.16	& 1.4	&	0.5	&	F1--3V&		\\
30	&	6 18 47.00	&	15 17 31.31	&	18.27	& 1.69	&	0.95	&	3.89	&	6.15	&	1.87	&	0.33		&	1.01		&	0.80&	0.23	& 1.6	&	1.2	&	F1--3V &  \\
31 &	6 18 47.42	&	15 17 12.06	&	18.38	& 1.91	&	0.72	&	2.74	&	4.93	&	2.21	&	0.19		&	1.12	&	0.85&	0.24	& 1.5	&	1.4		&	F2--5V & \\
32 &	6 18 45.89	&	15 16 34.89	&	19.21	&	1.05	&	3.13	&	-	&	-	&	2.23	&	0.34		&	1.05		&	0.46&	0.15	& 1.5	&	0.8	&	F2--3V\\
33 &	6 18 42.62	&	15 17 10.38	&	19.22	& 1.08	&	3.45&	2.13		&	4.04	&	2.94	&	-	&	1.59		&	0.52&	0.10	& 1.3	&	0.7	&	F7--8V\\
34 &	6 18 47.57	&	15 16 40.20	&	19.26	& 1.39	&	3.49	&	-	&	-	&	3.00	&	-	&	4.22		&	0.60&	0.15& 1.4	&	0.7	&	$\sim$G5V	\\
35 &	6 18 42.41	&	15 17 16.94	&	19.27	& 1.65	&	3.19&	-	&	-	&	2.99	&	-	&	6.26		&	0.60&	0.06	& 1.2	&	0.8	&	$\sim$G5V & fg$^f$\\
36 &	6 18 48.84	&	15 16 37.04	&	19.29	& 1.51	&	3.21&	-	&	-	&	2.93	&	-	&	4.81		&	0.49&	0.07	& 0.8	&	0.8	&	$\sim$G5V\\
37 &	6 18 46.96	&	15 17 23.71	&	19.34	&	0.99	&	4.50	&	-	&	-	&	2.78	&	-	&	2.01		&	0.34&	0.05& 1.4	&	0.4	&	$\sim$G5V\\
38 &	6 18 44.03	&	15 15 56.55	&	19.46	& 1.42	&	3.69&	-	&	-	&	2.71	&	-	&	2.79	&	0.34&	0.07	& 1.6	&	0.7		&	$\sim$G5V\\
39 &	6 18 46.43	&	15 16 21.13	&	19.59	& 1.18	&	3.82&	-	&	-	&	2.53	&	-	&	2.18		&	0.44&	0.14& 1.5	&	0.7	&	$\sim$G5V\\
\hline                                             
\multicolumn{17}{l}{Note. The rms error for the $V$ band magnitude is $\Delta V=0.67$~mag, for  $\Delta (V$--$I) = 0.38$~mag.} \\
\multicolumn{17}{l}{$^a$ Based on their individual extinctions listed in the second to last column of this table and assuming a distance modulus $\mu = 13.6$~mag.} \\
\multicolumn{17}{l}{$^b$ Determined via the Balmer emission lines decrement (Section \ref{balmer}).} \\
\multicolumn{17}{l}{$^c$ Determined via the spectral energy distribution (Section \ref{spt}). Most F- and G-type stars are either foreground stars or lie at the front side of the cluster, because of their lower} \\
\multicolumn{17}{l}{\phantom{$^c$} $E(B$--$V)_{SpT}$  (consistent with DIB measurements) and bright $V$-band magnitudes.} \\
\multicolumn{17}{l}{$^d$ Most studies assigned a spectral type ranging from late O to early B \citep{1994A&AS..103..503B,1996ApJ...458..653R,1994ASPC...62..134M}.} \\
\multicolumn{17}{l}{$^e$ Could be due to poor subtraction of the ambient nebular spectrum.} \\
\multicolumn{17}{l}{$^f$ Likely a foreground star, based on its radial velocity.} \\
\end{tabular}
\end{table}
\end{landscape}

\end{document}